\documentstyle[twoside,fleqn,npb,epsfig]{article}
\newcommand{\AmS}{{\protect\the\textfont2
  A\kern-.1667em\lower.5ex\hbox{M}\kern-.125emS}}

\title{News from the Muon (g-2) Experiment at BNL}

\author{M. Deile$^{11}$ for the Muon (g-2) Collaboration:\\
G.W. Bennett$^2$, B. Bousquet$^9$, H.N. Brown$^2$, G. Bunce$^2$, 
R.M. Carey$^1$, P. Cushman$^{9}$, G.T. Danby$^2$,
P.T. Debevec$^7$, M. Deile$^{11}$, H. Deng$^{11}$, W. Deninger$^7$, 
S.K. Dhawan$^{11}$, V.P. Druzhinin$^3$, L. Duong$^{9}$, E.~Efstathiadis$^1$, 
F.J.M. Farley$^{11}$,
G.V. Fedotovich$^3$, S. Giron$^{9}$, F. Gray$^7$, D. Grigoriev$^3$, 
M.~Gro\ss e-Perdekamp$^{11}$, 
A. Gro\ss mann$^6$, M.F. Hare$^1$, D.W. Hertzog$^7$, X. Huang$^1$, 
V.W. Hughes$^{11}$, M.~Iwasaki$^{10}$, 
K. Jungmann$^5$, D. Kawall$^{11}$, B.I. Khazin$^3$, J. Kindem$^{9}$,
F. Krienen$^1$, I. Kronkvist$^{9}$, A.~Lam$^1$, R. Larsen$^2$, Y.Y. Lee$^2$, 
I. Logashenko$^1$,
R. McNabb$^{9}$, W. Meng$^2$, J. Mi$^2$, J.P. Miller$^1$, W.M.~Morse$^2$, 
D. Nikas$^2$,
C. Onderwater$^7$, Y. Orlov$^4$, C.S. \"{O}zben$^2$, J. Paley$^1$, Q. Peng$^2$,
C. Polly$^7$, J.~Pretz$^{11}$,
R. Prigl$^2$, G.~zu Putlitz$^6$, T. Qian$^9$, S.I. Redin$^{3,11}$, 
O. Rind$^1$, B.L. Roberts$^1$,
N.M. Ryskulov$^3$, Y.K. Semertzidis$^2$, P. Shagin$^9$, Yu.M. Shatunov$^3$,
E. Sichtermann$^{11}$, E. Solodov$^3$, M. Sossong$^7$, A.~Steinmetz$^{11}$, 
L.R. Sulak$^1$,
A. Trofimov$^1$, D. Urner$^7$, P. von Walter$^6$, D. Warburton$^2$,
A. Yamamoto$^8$.\\
{\it $^1$Boston University, Boston, Massachusetts 02215, USA}.\\
{\it $^2$Brookhaven National Laboratory, Physics Dept., Upton, NY 11973, USA}.
\\
{\it $^3$Budker Institute of Nuclear Physics, Novosibirsk, Russia}.\\
{\it $^4$Newman Laboratory, Cornell University, Ithaca, NY 14853, USA}.\\
{\it $^5$Kernfysisch Versneller Instituut, Rijksuniversiteit Groningen, 9747
AA Groningen, The Netherlands}.\\
{\it $^6$Physikalisches Institut der Universit\"{a}t Heidelberg, 
69120 Heidelberg, Germany}.\\
{\it $^7$University of Illinois, Physics Dept., Urbana-Champaign, IL 61801, 
USA}.\\
{\it $^8$KEK, High Energy Accelerator Research Organization, Tsukuba, 
Ibaraki 305-0801, Japan}.\\
{\it $^{9}$University of Minnesota, Physics Dept., Minneapolis, 
MN 55455, USA}.\\
{\it $^{10}$Tokyo Institute of Technology, Tokyo, Japan}.\\
{\it $^{11}$Yale University, Physics Dept., New Haven, CT 06520,USA}.}
\begin{document}

\begin{abstract}
The magnetic moment anomaly $a_{\mu} = (g_{\mu} - 2) / 2$ of the positive muon 
has been measured
at the Brookhaven Alternating Gradient Synchrotron with an uncertainty of
0.7\,ppm. The new result, based on data taken in 2000, agrees well with 
previous measurements. Standard Model evaluations currently differ from the
experimental result by 1.6 to 3.0 standard deviations.
\end{abstract}

\maketitle

\section{Introduction}
According to Dirac's theory, the gyromagnetic ratio $g$ of a lepton
is exactly 2. Deviations from this prediction are
caused by radiative corrections to the lepton-photon vertex 
due to quantum field fluctuations.
The anomaly of the electron
is currently known at the level of 4\,ppb~\cite{dehmelt} and well in 
agreement with the Standard Model.
Since the contribution of heavy virtual particles to the anomaly 
$a = (g - 2) / 2$ is proportional to the square of the mass scale,
the sensitivity of the muon is enhanced by a factor 
$(m_{\mu}/m_e)^2 \approx 40000$ relative to the electron.
At the present level of precision, the muon anomaly $a_{\mu}$ probes
QED, weak and hadronic contributions.


\section{Experimental Setup and Data Analysis}
The general technique of the experiment at BNL is the same as that of the 
precursor experiment at CERN~\cite{CERN3}. Polarised muons are stored in a 
highly uniform
magnetic dipole field with electrostatic quadrupoles~\cite{quads} 
providing vertical
focussing. The muon spin precesses relative to the momentum vector with the
angular frequency
\begin{equation}
\label{eqn_precession}
\vec \omega_a = - {e \over m_\mu }\left[ a_{\mu} \vec B -
\left( a_{\mu}- {1 \over \gamma_\mu^2 - 1}\right)
\vec \beta \times \vec E \right],
\end{equation}
provided that $\vec \beta \cdot \vec B = 0$.
The dependence of $\omega_a$ on the electric field $\vec{E}$ (second term in
Eq.~(\ref{eqn_precession}))
is eliminated by storing the muons at the ``magic'' $\gamma_{\mu} = 29.3$, 
corresponding to a momentum $p_{\mu}$ = 3.094\,GeV/$c$.
In this case, $a_{\mu}$ is given by simultaneous measurement of $\omega_a$ 
and $\langle B \rangle$, the magnetic field averaged over the spatial
muon distribution in the storage region. As explained later, $\omega_a$ is
reflected by the rate of decay positrons above a certain energy threshold.
The magnetic field $B$ is measured in terms of
the free proton precession frequency $\omega_{\mathrm p}$ in this field using
nuclear magnetic resonance (NMR) probes. Then, $a_\mu$ can be expressed as
\begin{equation}
\label{eqn_amu}
a_\mu=\frac{ \omega_a/\omega_{\mathrm p} }{ \mu_\mu/\mu_{\mathrm p}
- \omega_a/\omega_{\mathrm p} } \: ,
\end{equation}
where $\mu_\mu/\mu_{\mathrm p} = (318\,334\,539 \pm 10) \times 10^{-8}$ 
is the ratio of the magnetic moments of the muon to the proton, which has been 
measured with a precision of 30\,ppb~\cite{liu}. 
To avoid experimenter biases, a ``blind analysis'' strategy is pursued,
separating the $\omega_{\mathrm a}$ and $\omega_{\mathrm p}$ analyses with
secret offsets which are only revealed when both analyses are complete and
internally consistent.

The beam used for the experiment originates at the Alternating Gradient 
Synchrotron (AGS) which every 2.5\,s delivers 
40--60\,$\times$\,10$^{12}$ protons at 24 GeV/$c$
onto a nickel target.
Each proton spill is composed of 12 bunches with a width of about 50 ns and a 
separation of~33\,ms.
Downstream of the target, pions at 3.1 GeV/$c$ are selected into a 72\,m long 
straight
beam line where about half of them decay into muons. Because of parity 
violation in the pion decay, the selection of forward-going muons leads to 
a polarisation of 
about 96\,\%. Muons at the magic momentum are selected and 
injected into the
storage ring through a hole in the yoke of the dipole magnet whose 1.45\,T
field is locally cancelled by
a DC super-conducting inflector magnet~\cite{inflect}. 
To move the muons onto the central orbit, a kick of about 11\,mrad
is given by a pulsed kicker magnet~\cite{kicker}.
The continuous superferric `C'-shaped storage ring magnet~\cite{danby}
is excited by superconducting coils.
The muon storage region has a 9\,cm diameter cross-section and a central 
radius of 7.112\,m, corresponding to
a cyclotron period of 149.2\,ns for muons at $\gamma = 29.3$.

A vertical air gap between pole and yoke decouples
the yoke and pole pieces, which are fabricated from high quality
steel, and allows the insertion of iron wedges to improve the field homogeneity
by compensating the quadrupole field components. 
The four edge shims, 5\,cm wide and about 3\,mm
high, are the main tool for reducing field variations over the beam 
cross-section. Surface coils glued to the pole pieces are used to
further reduce the inhomogeneity of the field. 

The field inside the storage region is mapped twice
a week using a hermetically sealed cable-car with a matrix of 17 NMR probes 
moving on rails in the vacuum beam pipe and measuring a transverse field map
about every 5\,mm. The probes
in the trolley are calibrated in place relative to a standard H$_{2}$O
probe for which the calibration ratio 
$\omega_{p}({\rm H}_{2}{\rm O})/\omega_{p}({\rm free})$ between the precession
frequencies of protons bound in water and free protons is known to 
50\,ppb~\cite{fei}. 

During data taking, an array of 375~NMR probes
embedded in the top and bottom plates of the vacuum chamber
is used to monitor magnetic field variations between trolley measurements 
and to stabilise the field with a feed-back
loop to the main magnet power supply~\cite{ralf}.

\begin{figure}[h!]
\center
\epsfig{file=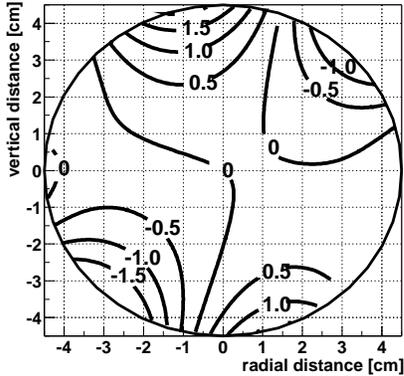,height=5cm}
\vspace*{-6mm}
\caption{\small Magnetic field map of the storage ring cross-section
averaged over azimuth. The contour lines have a distance of 0.5\,ppm.}
\label{fig_fieldmap}
\vspace*{-5mm}
\end{figure}

Two independent analyses determined $\langle \omega_{p} \rangle$, averaged
over the muon distribution. Their results agree within 0.05\,ppm.
The systematic uncertainties for $\langle \omega_{p}\rangle $ 
are summarised in Table~\ref{tab_omega_p}. 
As a final result, the value $\langle \omega_{p}\rangle /(2\pi) 
= 61\,791\,595(00)(15)$\,Hz (0.2\,ppm) was obtained.
The improvement from 0.4\,ppm
systematic error in 1999 to 0.24\,ppm in 2000 comes mainly 
from the better field homogeneity (Figure~\ref{fig_fieldmap}) which was 
achieved by replacing the old inflector whose super-conducting fringe-field 
shield had a flux leak. 

\begin{table}[h!]
\setlength{\tabcolsep}{1.5pc}
\newlength{\digitwidth} \settowidth{\digitwidth}{\rm 0}
\catcode`?=\active \def?{\kern\digitwidth}
\caption{Systematic uncertainties for the $\omega_{p}$ analysis.}
\label{tab_omega_p}
\begin{tabular*}{75mm}{@{}l@{\extracolsep{\fill}}c}
\hline
Source of errors & Size [ppm] \\
\hline
Absolute calibration of standard probe & 0.05 \\
Calibration of trolley probes          & 0.15 \\
Trolley measurements of central        &      \\
azimuthal average field                & 0.10 \\
Interpolation with fixed probes        & 0.10 \\
Uncertainty from muon distribution     & 0.03 \\
Others$^{\dag}$                        & 0.10 \\
\hline
Total systematic error on $\omega_{p}$ & 0.24 \\
\hline
\multicolumn{2}{@{}p{75mm}}{\small $^{\dag}$ higher multipoles, trolley 
temperature 
and voltage response, eddy currents from kickers, and time-varying stray 
fields.}
\end{tabular*}
\vspace*{-5mm}
\end{table}

The decay positrons from $\mu^+ \rightarrow \mathrm{e^+ \nu_{e} \bar
\nu_{\mu}}$ range in energy from 
0 to 3.1\,GeV, and are detected with 24 lead/scintillating-fiber 
calorimeters~\cite{sedykh}
placed symmetrically around the inside of the storage 
ring. The arrival times and energies of the positrons are determined from the 
calorimeter pulses whose full shapes are sampled by a 400\,MHz waveform 
digitiser (WFD). A laser and light-emitting-diode (LED) system is
used to monitor potential time and gain shifts.

\begin{figure}[h!]
\center
\small
\vspace*{-5mm}
\epsfig{file=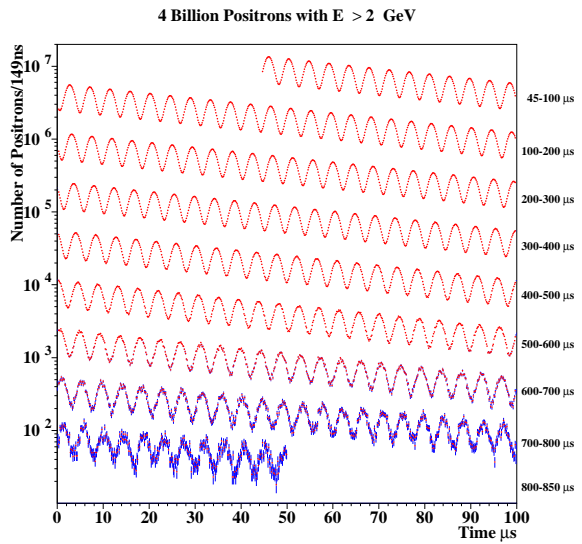,width=75mm}
\vspace*{-10mm}
\caption{\small Positron time spectrum from the 2000 data set with all 
detectors combined.}
\label{fig_wiggles}
\vspace*{-5mm}
\end{figure}

Because of parity violation in the weak muon decay, in the muon rest frame 
positrons are preferentially emitted along the muon spin direction. 
Since in the lab frame forward positrons are boosted to high energies,
the muon spin precession frequency modulates
the decay positron count rate $N(t)$ if a lower energy threshold is applied:
\begin{equation}
\label{eqn_5parfit}
N(t) = N_0\,{\rm e}^{-\frac{t}{\gamma\tau}} 
\left[1 + A \cos \left(\omega_a t+\phi_{a}\right)\right] \: ,
\end{equation}
\noindent
where $\gamma\tau = 64.4\,\mu$s is the dilated muon lifetime. For an
energy threshold of 2\,GeV, the asymmetry $A$ is about 0.4.
Figure~\ref{fig_wiggles} shows the sum of the decay positron time spectra 
observed by all
detectors within a time range of 805\,$\mu$s or roughly 12 muon lifetimes. 
The total number of positrons recorded later than 45\,$\mu$s after injection
is about $4 \times 10^{9}$.
Error bars are drawn on all points but are only visible at very late times 
because 
of the huge statistics of up to $10^{7}$ entries per 149\,ns time bin. Given 
this high number of events, the simple parametrisation in 
Eq.~(\ref{eqn_5parfit}) proved not to be adequate for fitting the spectrum
which is affected by several significant perturbations.

We shall concentrate here on the leading systematic effect caused by 
coherent betatron oscillations -- 
oscillations of the beam as a whole. Since the inflector aperture is smaller 
than the storage ring aperture, the phase space for betatron oscillations is
not filled, which leads to a radial modulation of beam width and centroid.
Looking from fixed detector positions, these oscillations have a frequency
of 466\,kHz, approximately given by
$f_{\rm CBO} \approx (1 - \sqrt{1 - n}) f_{c}$ where 
$f_{c} = 6.7$\,MHz is the 
cyclotron frequency and $n = 0.136$ is the field index adjusted by the
electric quadrupole voltage~\cite{quads}. 
The value of $n$ was chosen far away from 
physical resonance conditions which could lead to increased muon losses or 
spin flips. Since both the calorimeter acceptance and the 
energy distribution of the detected positrons depend on the radial position 
of the muon decay, the parameters $N_{0}$, $A$ and $\phi_{a}$ in 
Eq.~(\ref{eqn_5parfit}) receive a time modulation, e.g. 
$N_{0}(t) = N_{0} [1 + g_{N}(t) \cos(2 \pi f_{\rm CBO} t + \phi_{\rm CBO})]$,
where $g_{N}(t)$ describes the CBO decay due to the muon momentum spread and
higher $\vec{E}$ and $\vec{B}$ field multipoles.
The time constant is typically 100\,$\mu$s.
The modulation $N_{0}(t)$ -- with $g_{N}(0) \approx 1$\,\% -- 
was already observed and included in the analysis of the 1999 data set.
Its parameters do not correlate strongly with $\omega_{a}$. 
The modulations $A(t)$ and $\phi_{a}(t)$ are smaller
-- 0.1\,\% and 1\,mrad at $t=0$ respectively -- and were only discovered 
with the higher
statistics of the 2000 data set. Their dangerous effect on the time spectrum 
is the creation of interference terms with the frequency 
$f_{\rm CBO} - \omega_{a} / (2\pi)$ which for our choice of $n$ is very
close to $\omega_{a}$. This parametric resonance can produce 
shifts in $\omega_{a}$ of up to 4\,ppm if individual 
detector time spectra are fitted without including $A(t)$ and $\phi_{a}(t)$.
In the sum of all detector spectra, the shifts largely cancel 
thanks to the circular symmetry of the (g-2) ring. 
Remaining effects are accounted for by including $A(t)$ into the fit
and assigning a systematic error for the impact of $\phi_{a}(t)$. Including
$\phi_{a}(t)$ into the fit turned out to be more difficult because this
term induces strong correlations between detector gain changes and 
$\omega_{a}$, entailing further shifts in $\omega_{a}$.

Other perturbations were treated
like in earlier data sets~\cite{prl2001}. Pulse pileup effects were removed
by statistically superimposing recorded pulses from the data themselves and 
thus constructing an artificial pileup spectrum which was then subtracted from
the untreated spectrum. The effects of beam debunching were eliminated by
randomizing the start time of each fill over one cyclotron period.
Muon losses were taken into account by multiplying the function
in Eq.~(\ref{eqn_5parfit}) with an extra loss term. 
AGS background due
to erroneous proton extraction during the muon storage period, which 
contributed 0.1\,ppm uncertainty to the 1999 result, was largely eliminated 
by installing a sweeper magnet in the beam line.

Four independent $\omega_{a}$ analyses with different approaches to
take systematic effects into account were completed.
Their results agreed within 
0.4\,ppm -- as compared to statistically allowed variations of 0.5\,ppm --
and were combined to 
$\omega_{a}/(2\pi) = 229\,074.11(14)(7)$\,Hz (0.7\,ppm). This number contains a
correction of +0.76(3)\,ppm for residual effects of the electric field on
muons with $\gamma \ne 29.3$ and for deviations from Eq.~(\ref{eqn_precession})
due to vertical beam oscillations (i.e. $\vec \beta \cdot \vec B \ne 0$). 
The combined systematic
errors listed in Table~\ref{tab_omega_a} account for the
correlations between the results from the individual analyses.

\begin{table}[h!]
\vspace*{-5mm}
\setlength{\tabcolsep}{1.5pc}
\settowidth{\digitwidth}{\rm 0}
\catcode`?=\active \def?{\kern\digitwidth}
\caption{Systematic uncertainties for the $\omega_{a}$ analysis.}
\label{tab_omega_a}
\begin{tabular*}{75mm}{@{}l@{\extracolsep{\fill}}c}
\hline
Source of errors & Size [ppm] \\
\hline
Coherent betatron oscillations         & 0.21 \\
Pileup                                 & 0.13 \\
Detector gain changes                  & 0.13 \\
Lost muons                             & 0.10 \\
Binning and fitting procedure          & 0.06 \\
Others$^{\dag}$                        & 0.06 \\
\hline
Total systematic error on $\omega_{a}$ & 0.31 \\
\hline
\multicolumn{2}{@{}p{75mm}}{\small $^{\dag}$ AGS background, timing shifts, 
E field and vertical oscillations, beam debunching and randomisation.}
\end{tabular*}
\vspace*{-5mm}
\end{table}


\section{Result from the Data Set of 2000 and Comparison with Theory}
\label{sec_result}
\begin{figure}[h!]
\center
\small
\vspace*{-5mm}
\epsfig{file=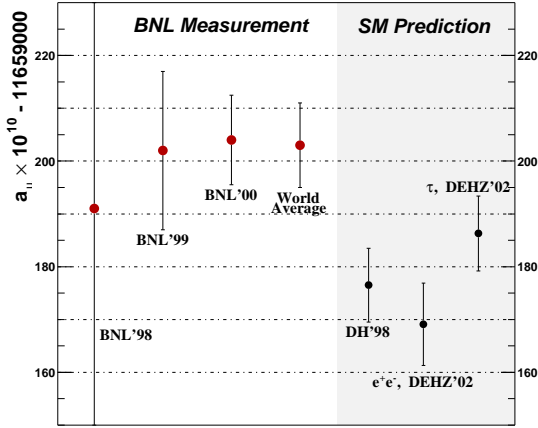,width=75mm}
\vspace*{-1.5cm}
\caption{\small Recent experimental and theoretical values of $a_{\mu}$.  
The labels for the theoretical predictions represent the evaluation of
$a_{\mu}({\rm had, 1})$ (see text).}
\label{fig_result}
\vspace*{-5mm}
\end{figure}

After completion of the $\omega_{a}$ and $\omega_{p}$ analyses, $a_{\mu}$ was
calculated according to Eq.~(\ref{eqn_amu}). The result is
$a_{\mu^{+}} = 11\,659\,204\,(7)(5) \times 10^{-10}$ (0.7\,ppm)~\cite{prl2002}.
 It agrees
well with older measurements (Figure~\ref{fig_result}). The new experimental
world average,
$a_{\mu^{+}} = 11\,659\,203\,(8) \times 10^{-10}$ (0.7\,ppm), is dominated
by the new result which has about half the uncertainty of previous 
measurements.

The Standard Model prediction for $a_{\mu}$~\cite{dehz02} can be written as
\begin{equation}
a_{\mu}({\rm SM}) = a_{\mu}({\rm QED}) + a_{\mu}({\rm weak}) + 
a_{\mu}({\rm hadronic})
\end{equation}
with $a_{\mu}({\rm QED}) = 11\,658\,470.57(0.29) \times 10^{-10}$ 
and $a_{\mu}({\rm weak}) = 15.1(0.4) \times 10^{-10}$.
The hadronic contribution cannot be calculated from first principles at 
this time because it is dominated by low-energy interactions.
The first-order hadronic vacuum polarisation contribution, 
$a_{\mu}({\rm had, 1})$, can be determined from measured $e^{+}e^{-}$ 
annihilation cross-sections over all energies using a dispersion relation.
It can also be related to hadronic $\tau$ decay. These calculations are
still under theoretical 
investigation. Higher-order contributions are given by
$a_{\mu}({\rm had, 2}) = -10.0(0.6) \times 10^{-10}$
and $a_{\mu}$(had, light-by-light) 
= +8.6(3.2) $\times 10^{-10}$. 
For the latter, a sign error was recently corrected.
Figure~\ref{fig_result} shows the theoretical predictions
for $a_{\mu}$ using recent evaluations 
of $a_{\mu}({\rm had, 1})$. DH'98~\cite{dh98} is the value used for the 
comparison in our PRL~\cite{prl2002}. In this evaluation,
$a_{\mu}({\rm had, 1})$ uses data from both $e^{+}e^{-}$ annihilation 
and $\tau$ decay. 
Later, new $e^{+}e^{-}$ 
data from Novosibirsk~\cite{akhmetshin} gave rise to a new calculation of
$a_{\mu}({\rm had, 1})$ which does not agree any more with the 
value obtained from $\tau$ decay data (DEHZ'02:~\cite{dehz02}).
Using the $e^{+}e^{-}$-based result,
$a_{\mu}({\rm had, 1}) = 684.7(7.0)\times 10^{-10}$, one obtains a total
theory prediction $a_{\mu}({\rm SM}) = 11\,659\,169.1\,(7.8) \times 10^{-10}$ 
(0.7\,ppm). Using the $\tau$-based result,
$a_{\mu}({\rm had, 1}) = 701.9(6.1)\times 10^{-10}$, one obtains a total
theory prediction $a_{\mu}({\rm SM}) = 11\,659\,186.3\,(7.1) \times 10^{-10}$ 
(0.6\,ppm). The deviations of the two evaluations from the experimental result
correspond to 3.0 and 1.6 standard deviations, respectively. Hence, no 
unambiguous statement about new physics can be made at present.

\section{Outlook}
In the year 2001, the experiment was performed with negative muons and
with different field indices $n$ moving the CBO frequency away from the
parametric resonance.
The resulting data set of about $3 \times 10^{9}$ electrons is currently being 
analysed. It will 
provide a test of CPT invariance and -- if CPT holds --
an improved combined value of $a_{\mu^{\pm}}$. However, in order to achieve
the design goal of 0.35\,ppm statistical uncertainty, additional 
$6 \times 10^{9}$ events are needed.
A new run is planned, but at present funding is uncertain.

\section*{Acknowledgements}
This work was supported by the U.S. Department of Energy, the U.S. 
National Science Foundation, the German Bundesminister f\"{u}r Bildung und
Forschung, the Russian Ministry of Science, and the US-Japan Agreement in 
High Energy Physics. Mario Deile acknowledges support by the Alexander von
Humboldt Foundation.

\end{document}